\documentclass[preprint,showpacs,preprintnumbers,amsmath,amssymb,prb,superscriptaddress]{revtex4}

\usepackage[letterpaper,left=0.8in,right=0.8in,top=1.00in,bottom=0.8in]{geometry}
\usepackage{graphicx}
\usepackage{dcolumn}
\usepackage{bm}
\usepackage{graphicx}
\usepackage{epsfig}
\usepackage{floatflt}
\usepackage{amsmath}
\input amssym.def
\input amssym.tex

\def\alloy#1,#2,#3{#1$_{1-x}$#2$_{x}$#3}

\def\cro2{CrO$_2$}

\def\ftn[#1]{\footnotemark[#1]}

\def\alloy#1,#2,#3{#1$_{1-x}$#2$_{x}$#3}

\def\cro2{CrO$_2$}

\def\H0{H^0}

\newcommand{\bfr}{{\bf r}}
\newcommand{\bfrp}{{\bf r}^\prime}
\newcommand{\bfq}{{\bf q}}

\newcommand{\qsgw}{QS\emph{GW}}

\newcommand{\vtot}{\delta V}
\newcommand{\vext}{\delta V^{\rm ext}}

\begin{document}

\title{First-principles Theory of Nonlocal Screening in Graphene}

\author{M. van Schilfgaarde}
\affiliation{School of Materials, Arizona State University, Tempe AZ}
\author{M. I. Katsnelson}
\affiliation{Radboud University Nijmegen, Institute for Molecules
and Materials, 6525AJ Nijmegen, The Netherlands}

\begin{abstract}
Using the quasiparticle self-consistent \emph{GW} (\qsgw) and local-density
(LD) approximations, we calculate the $q$-dependent static dielectric
function, and derive an effective 2D dielectric function corresponding to
screening of point charges.  In the $q${}$\to$0 limit, the 2D function is
found to scale approximately as the square root of the macroscopic
dielectric function.  Its value is $\simeq$4, a factor approximately 1.5
larger than predictions of Dirac model.  Both kinds of dielectric functions
depend strongly on $q$, in contrast with the Dirac model.  The \qsgw\
approximation is shown to describe QP levels very well, with small
systematic errors analogous to bulk $sp$ semiconductors.  Local-field
effects are rather more important in graphene than in bulk semiconductors.
\end{abstract}


\maketitle

\setcounter{page}{1}

Graphene is a first truly two-dimensional (2D) crystal, with unique
electronic and structural properties (for review, see
Refs.~\onlinecite{r1,r2,r3,r4,r5}).  Screening of electron-electron and
electron-impurity interactions in graphene is an important theoretical
issue crucial for both many-body effects in electronic structure
\cite{ggw94} and for transport properties, especially, for electron
scattering by charge impurities \cite{r5,ando_jpsj_2006}.  There are
numerous
works\cite{ando_jpsj_2006,guinea_njp_2006,katsnelson_prb_2006,shytov_prl_2007,fogler_prb_2007,polini_prb_2008,rossi_prl_2008,brey_prb_2009,fogler_prl_2009}
treating this issue within the two-band Dirac model.  But the Dirac model
does not take into account the many other bands involved, which can include
van Hove singularities in electron density of states \cite{r3} that may
possibly be very essential, specifically for screening
\cite{KT_anom,KT_anomb}.  Here we develop a definition for an effective 2D
dielectric function in an \emph{ab initio} context, and calculate it within
the Quasiparticle self-consistent \emph{GW} (\qsgw) and local-density (LD)
approximations.  The former takes into account many-body effects beyond the
density functional GGA or LDA schemes essential for correct description of
excited states and thus screening effects \cite{ferdi98,Onida02}.

There are several \emph{GW} calculations for graphene
\cite{GW_graphene1,Yang09,GW_graphene3}, where $G$ and the screened
Coulomb interaction $W$ are computed from the LDA.  They all predict a
notable (20-40\%) increase of the Fermi velocity $v_F$ at the Dirac point
$K$ relative to the LDA(GGA) value, with $v_F$ between 1.1 and
1.2$\cdot10^6$ m/s, in very good agreement with experiment
\cite{r1,r2,r3,r4}.  The dielectric function and optical conductivity as a
function of frequency $\omega$ for zero wave vector ${\bf q}$=0 was also
calculated in Refs.  \onlinecite{GW_graphene1,Yang09}.  Here we focus
on the static dielectric function ($\omega$=0) as a function of $q$. This
quantity is relevant for calculations of resistivity via charge impurities
\cite{r5,ando_jpsj_2006}, as well as for the problem of supercritical
Coulomb centers \cite{shytov_prl_2007,shytov2,pereira_prl_2007} and
possible exciton instabilities \cite{guinea_prb_2010}.


The inverse dielectric function $\epsilon^{-1}(\bfr,\bfrp,\omega)$
relates the change in total potential $\vtot$ to an external
perturbing potential $\vext$ as \cite{ferdi98,Onida02}
\begin{eqnarray}
\vtot(\bfr,\omega)=\int{d\bfrp\epsilon^{-1}(\bfr,\bfrp,\omega)\vext({\bfrp},\omega)}.
\label{eq:defeps}
\end{eqnarray}
$\epsilon^{-1}$ is obtained from a convolution of the polarization
operator $\Pi$ and the bare Coulomb interaction $v$ as
\begin{eqnarray*}
\epsilon^{-1} = \left( {1 - v\Pi } \right)^{-1}.
\label{eq:defoeps}
\end{eqnarray*}


In a system with translation symmetry, $\epsilon^{-1}$, $\Pi$, and $v$
can be expanded in Bloch functions $\{B_I^{\bf{q}}({\bfr})\}$,
e.g.
\begin{eqnarray}
\epsilon^{-1} ({\bf{r}},{\bf{r'}},\omega ) = \sum\nolimits_{{\bf{q}}IJ} {B_I^{\bf{q}} ({\bf{r}})\epsilon^{-1}_{IJ}({\bf{q}},\omega )} {B_J^{\bf{q}}}^* ({\bf{r'}})
\label{eq:epsbasis}
\end{eqnarray}
The most common choice of $\{B_I^{\bf{q}}({\bfr})\}$ are plane waves,
\begin{eqnarray}
B_I^{\bf{q}}({\bfr})\to B_{\bf{G}}^{\bf{q}} ({\bf{r}}) = \exp
(i({\bf{q}} + {\bf{G}})\cdot{\bf{r}}), \label{eq:pwbasis}
\end{eqnarray}
${\bf G}$ being reciprocal lattice vectors.

Quantities of interest are coarse-grained averages of
$\epsilon^{-1}_{\bf{GG}'}({\bf{q}},\omega)$. The ``macroscopic'' response to a plane
wave perturbation is~\cite{Onida02}
\begin{eqnarray}
\epsilon_M({\bf{q}},\omega) = \left[ {\epsilon
_{{\bf{G}=0,{G'}=0}}^{ - 1} ({\bf{q}},\omega )} \right]^{-1}
\label{eq:epsm}
\end{eqnarray}
The matrix structure of $\epsilon^{-1}$ with ${\bf G} \neq {\bf{G}^\prime}$
reflects local field effects in terms of classical electrodynamics. The
quantity $\epsilon_M({\bf q})$ is commonly approximated by just
$\epsilon({\bf q})$; that is, the Umklapp processes, or local field effects
are neglected. This is not such a bad approximation in $sp$ semiconductors
but as we show here, it is a rather poor approximation in graphene.
$\epsilon_M({\bf q})$ corresponds to screening potential $\vext$ with a
single Fourier component ${\bf q}$.  Selecting ${\bf{G}}$=${\bf{G'}}$=0
averages $\epsilon^{-1}$ over the unit cell, restricting the spatial
variation to the envelope $\exp(i{\bf q}\cdot{\bf r})$.  While $\epsilon_M$
is a quantity of relevance to some experiments, perhaps the most relevant
is screening of a point charge in the graphene sheet, which governs e.g.,
scattering from impurities.

As graphene is a 2D system, we need to consider how the impurity
potential $v(q)=4\pi/q^2$ is screened in the sheet. The
(statically) screened potential from a point charge at the origin
may be written in cylindrical coordinates
$\bf{r}$=$(\rho,z,\theta)$ and $\bf{q}$=$(\bar{q},q_{z},\theta_q)$ as
\begin{eqnarray}
 W(\rho ,z) &=& \frac{1}{{2\pi }}\int_0^\infty  {d\bar{q}\,\bar{q}J_0 (\bar{q}\rho )\,W^{2{\rm{D}}} (\bar{q},z)}  \\
 W^{2{\rm{D}}} (\bar{q},z) &=& 4\int_0^\infty  {dq_z e^{iq_z z} \frac{{\epsilon ^{ - 1} (\bar{q},q_z )}}{{q_z^2  + \bar{q}^2 }}}
\label{eq:wcylindrical}
\end{eqnarray}
Thanks to graphene's hexagonal symmetry,
$\epsilon^{-1}$ does not depend on $\theta_q$ for small $\bar{q}$.
$W^{2{\rm{D}}}(\bar{q},z)$ is the 2D (Hankel) transform of $W({\bf{r}})$,
the analog of the 3D transform $W({\bf{q}})$=$\epsilon^{-1}({\bf{q}})v({\bf{q}})$.
In the absence of screening $\epsilon^{-1}=1$ and $W^{2{\rm{D}}}(\bar{q},z)$
reduces to the bare coulomb interaction $v^{2{\rm{D}}}(\bar
q,z)$:
\begin{eqnarray}
v^{2{\rm{D}}} (\bar{q},z) =  {4\int_0^\infty  {dq_z e^{iq_z z} \frac{1}{{q_z^2  + \bar{q}^2 }}} } = \frac{{2\pi }}{{\bar{q}}}e^{ - \bar{q}z}
\label{eq:vcylindrical}
\end{eqnarray}
An appropriate definition of an effective 2D dielectric function is then
\begin{eqnarray}
\epsilon^{2{\rm{D}}}(\bar{q},z) = v^{2{\rm{D}}} (\bar{q},z) / W^{2{\rm{D}}} (\bar{q},z)
\label{eq:eps2d}
\end{eqnarray}
Graphene wave functions have some extent in $z$ which must be
integrated over to obtain a scattering matrix element.  But the largest
contribution originates from $z$=0, so $W^{2{\rm{D}}}(\bar{q},0)$ is a
reasonable estimate for the scattering potential.  This is
particularly so for small $\bar{q}$ of primary interest here.

In practice we carry calculations in a periodic array of graphene sheets in
the $xy$ plane, spaced by a distance large enough that the sheets interact
negligibly.  To calculate $\epsilon^{-1}_{{\bf G}={\bf
G'}=0}({\bf{q}},\omega)$ we adopt the all-electron, augmented wave
implementation that was developed for the quasiparticle self-consistent
$GW$ (\qsgw) approximation, described in Ref.~\cite{Kotani07}.  It makes no
pseudo- or shape- approximation to the potential, and does not use PWs
(Eq.~\ref{eq:pwbasis}) for the product basis $\{B\}$, but a mixed basis
consisting of products of augmented functions in augmentation spheres, and
plane waves in the interstitial region.  The all-electron implementation
enables us to properly treat core states.
We calculate $\epsilon^{-1}({\bf{q}},\omega)$ in the random phase
approximation, using Bloch functions for eigenstates\cite{ferdi98}. These
are obtained from single-particle eigenfunctions $\Psi_{{\bf{}k}n}$ and
eigenvalues $\epsilon_{{\bf k}n}$ in both the LDA and \qsgw\
approximations.  In both cases the generalized LMTO method is
used~\cite{mark06adeq,basis}.

\qsgw\ has been shown to be an excellent predictor of materials properties
for many classes of compounds composed of elements throughout the Periodic
Table, with unprecedented ability to consistently and reliably predict
materials properties over a wide range of
materials~\cite{Kotani07,Faleev04,mark06qsgw,Chantis06a,Chantis07a}.
Nevertheless there are small, systematic errors: in particular bandgaps in
insulators such as GaAs, SrTiO$_{3}$ and NiO, are systematically
overestimated.  Its origin can be traced to a large extent to the RPA
approximation to the polarizability, $\Pi^{\rm{RPA}}$=$iGG$.  The RPA
bubble diagrams omit electron-hole interactions in their intermediate
states.  Short-range attractive (electron-hole) interactions induce
redshifts in Im$\,\epsilon(\bfq,\omega)$ at energies well above the
fundamental bandgap; see e.g. Fig. 6 in Ref.\cite{Kotani07}. That ladder
diagrams are sufficient to remedy most of the important errors in
$\Pi^{\rm{RPA}}$ was demonstrated rather convincingly in Cu$_{2}$O, by
Bruneval et al.~\cite{Bruneval06b}.  Moreover Shishkin et
al~\cite{Shishkin07} incorporated these ladder diagrams in an approximate
way for several $sp$ semiconductors, and established that they do in fact
largely ameliorate the gap errors.  Yang et al.  investigated the effect of
ladder diagrams in graphene and graphite, and showed that in a manner very
analogous to ordinary semiconductors, these diagrams induce a redshift in
the peak of Im$\,\epsilon^{{\rm{RPA}}}(\bfq,\omega)$ near
5~eV,~\cite{Yang09} of $\sim$0.6~eV.  They found a strong similarity with
conventional semiconductors, namely that the redshift from ladder diagrams
approximately cancels the error in the LDA joint density of states.

\begin{table}[ht]
\caption{ Energy gap $E_{G}$ and valence bandwidth $\Gamma_{1v}$ in diamond
   (eV); Fermi velocities $v_{F}$ in graphite and graphene ($10^6$m/sec).
   There is a significant renormalization of the bandgap from the
   electron-phonon interaction in diamond, estimated to be 370~meV
   \cite{Cardona05}.  Thus \qsgw\ overestimates $E_{G}$ by a slightly
   smaller amount than in other semiconductors, and the scaling of $\Sigma$
   as described in the text results in a slightly underestimated gap.  The
   electron-phonon interaction also reduces the Fermi velocity in graphene,
   estimated to be 4 to 8\% in an LDA-linear response calculation
   \cite{Park07}.  The calculated Fermi velocities should be reduced by
   this much when comparing to experiment.  $v_{F}$ calculated by \qsgw\ is
   slightly overestimated, for much the same reason semiconductor gaps are
   overestimated. $v_{F}$ calculated from the scaled-$\Sigma$ potential, is
   slightly larger than $v_{F}$ calculated LDA-based $GW$, i.e. $G^{\rm
   LDA}W^{\rm LDA}$\cite{Reining08}, just as semiconductor bandgaps are
   slightly larger.  When renormalized by the electron-phonon interaction,
   $v_{F}$ agrees very well with the measured value\cite{Zhang05}.}
\label{tab:diamond}
\begin{ruledtabular}
\begin{tabular}{crrrr}
                             & LDA     & \qsgw      & scaled $\Sigma$ & Expt           \\
  $E_{G}$, diamond           & $4.09$  & $5.93$     & $5.56$          & $5.50$         \\
  $\Gamma_{1v}$, diamond     & $21.3$  & $23.1$     & $22.7$          & $23.0\pm 0.2$\ftn[1] \\
  $\Gamma_{1v}$, graphene    & $19.4$  & $22.9$     & $22.2$          &               \\
  $v_{F}$(H), graphite       & $0.77$  & $0.99$     & $0.94$          & $0.91\pm 0.15$ \\
  $v_{F}$(K), graphene       & $0.82$  & $1.29$     & $1.20$          & $1.1$
\end{tabular}
\footnotetext[1]{Ref.~\onlinecite{Jimenez97}}
\end{ruledtabular}
\end{table}

\begin{figure}[h]
\includegraphics[height=5.5cm]{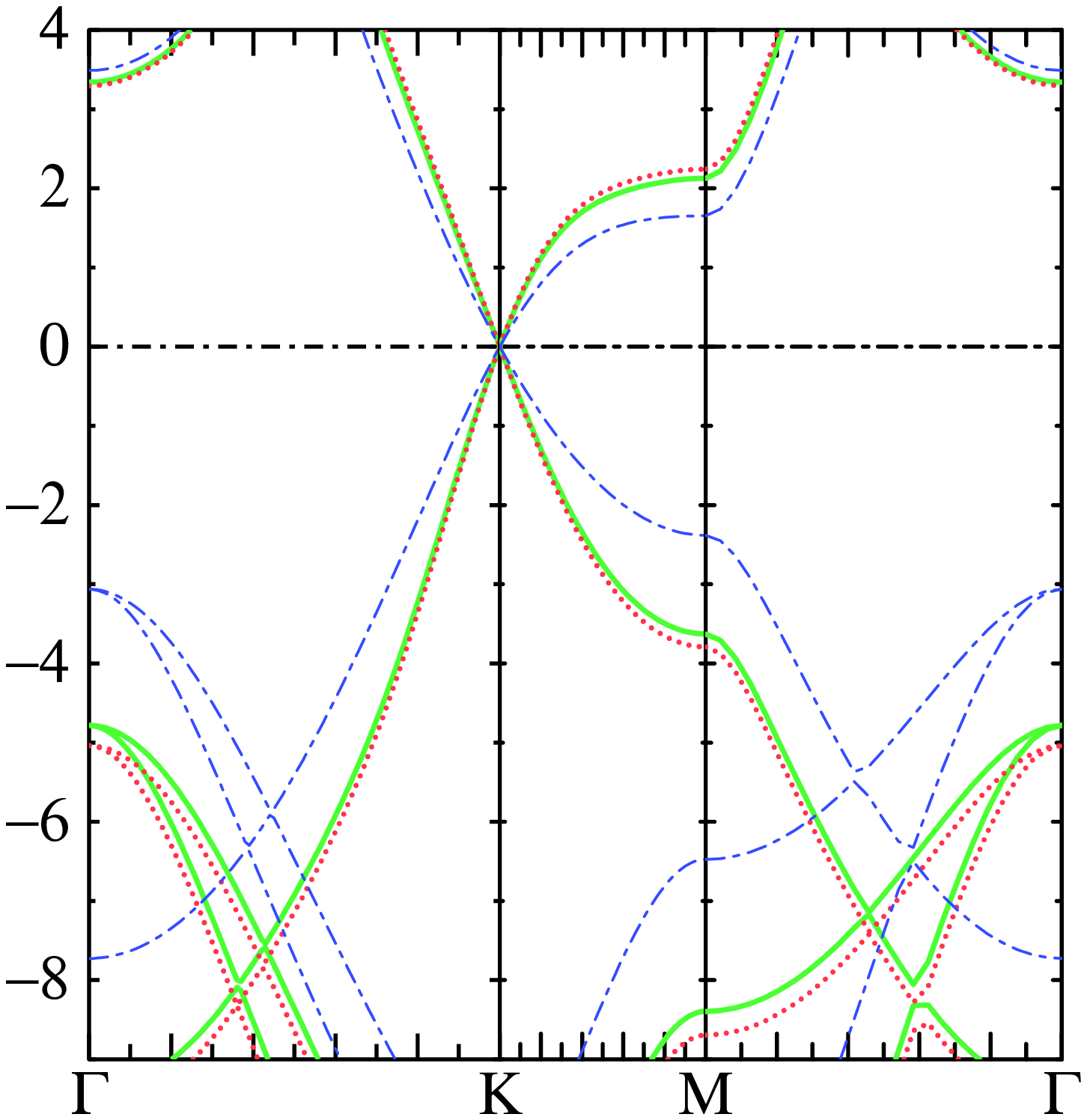}
\caption {\qsgw\ bands of graphene (dotted red lines), compared to LDA
results (dashed blue lines) and \qsgw\ results with $\Sigma$ scaled by 0.8
(solid green lines) described in the text. The linear dispersion near K (or
H, in graphite) is significantly larger in the \qsgw\ case.  Differences are
quantified in Table~\ref{tab:diamond}.  The lowest lying unbound state can
be seen as a parabolic band starting at $\Gamma$ near 3.5~eV.  It
corresponds to the work function.  LDA and \qsgw\ work functions are very
similar, consistent with the observation that LDA predicts work functions
rather well in many systems.}
\label{fig:bands}
\end{figure}

A redshift in the peak of Im$\epsilon(\omega)$ increases the static
dielectric constant $\epsilon_{\infty}$, as can be readily seen through the
Kramers-Kronig relations.  Remarkably, $\epsilon_{\infty}$ as calculated by
the RPA in \qsgw, is underestimated by a nearly \emph{universal factor} of
0.8, for many kinds of insulators and semiconductors~\cite{mark06qsgw},
including transition metal oxides such as NiO \cite{Kotani07}, CeO$_{2}$,
and \emph{sp} semiconductors~\cite{Shishkin07}.  (This error is often
approximately canceled in the LDA, fortuitously.  As Yang et al. noted, the
cancellation seems to apply to graphene in a manner similar to ordinary
semiconductors.)  Because $\epsilon$ is systematically
\emph{underestimated}, $W=\epsilon^{-1} v$ and $\Sigma=-iGW$ are
systematically \emph{overestimated}; therefore QP excitation energies are
also systematically overestimated.  We have found that simply scaling by
0.8 (the nearly universal ratio $\epsilon^{{\rm
QS}GW}_{\infty}/\epsilon^{\rm expt}_{\infty}$) largely eliminates
discrepancies between \qsgw\ and measured QP levels in a wide range of
$spd$ systems, including all zincblende semiconductors, and many other
kinds of insulators.  For graphene, we find that the \qsgw\ macroscopic
($q${}$\to$0) dielectric constant was found to be 80\% of the LDA one,
consistent with the universal pattern in bulk insulators noted above.  The
many points of consistency with 3D behavior, both in the \qsgw\ QP levels
and the dielectric response suggest that \qsgw\ will exhibit the same
reliable description of the 2D graphene system, with similar systematic
errors.  To confirm this, some band parameters for three pure (undoped)
carbon compounds calculated by \qsgw\ and \qsgw\ with $\Sigma$ scaled by
0.8 are shown in Table~\ref{tab:diamond}.  Scaling \qsgw\ has a minor
effect on the quasiparticle levels: e.g. it reduces $v_F$ by 7\%.  As
Table~\ref{tab:diamond} shows, $v_F$ falls in very close agreement with
experiment when $\Sigma$ is scaled and the electron-phonon interaction is
taken into account, consistent with agreement in gaps in the bulk
insulators.  Even though the \qsgw\ and LDA work functions are similar
(Fig.~\ref{fig:bands}), the valence band is significantly widened relative
to LDA,~\cite{Jimenez97} more so in graphene than in diamond.

Careful checks for convergence were made in various parameters.  To check
supercell artifacts, a ``small'' 3D unit cell with the graphene planes
repeated at a spacing equivalent to 4 atomic layers of graphite (25 a.u.)
was compared against a ``large'' cell, with graphene planes spaced at 8
layers.  The bands from $-\infty$ to $E_F$+5~eV were found to be a very
similar, with a slight increase in $v_F$ (1.23$\to$1.29 $\cdot10^6$
m/s). $k$ convergence in the construction of $\Sigma$ was monitored by
comparing QP levels generated on a 6$\times$6$\times$2 $k$ mesh to a
9$\times$9$\times$2 mesh.  QP levels were nearly identical:
$v_F$ differed by $<$1\% in the both the small and large 3D cells.

$\epsilon_{00}^{-1}(\bfq,\omega)$ must be integrated with a fine $k$ mesh.
To deal with the delicate ${\bf{q}}${}$\to$0 limit, we calculated
$\epsilon^{-1}$ integrating on a standard $k$ mesh including $\Gamma$, and
an offset mesh (Eqns. 47 and 52 in Ref.~\onlinecite{Kotani07}), and
averaged them.  We present data for averaged 18$\times$18$\times$4 meshes.
Calculations without local fields were also performed for a pair of
24$\times$24$\times$4 meshes.
$\epsilon(\bfq_{||},q_{z}$=0$,\omega$=0) calculated by 18- and 24-
(averaged) mesh integrations were essentially indistinguishable for
$q${}$>$0.1$\times{}2\pi/a$, and differed by a few percent for
$q${}$>$0.02$\times2\pi/a$.

\begin{figure}[h]
\includegraphics[height=3.5cm]{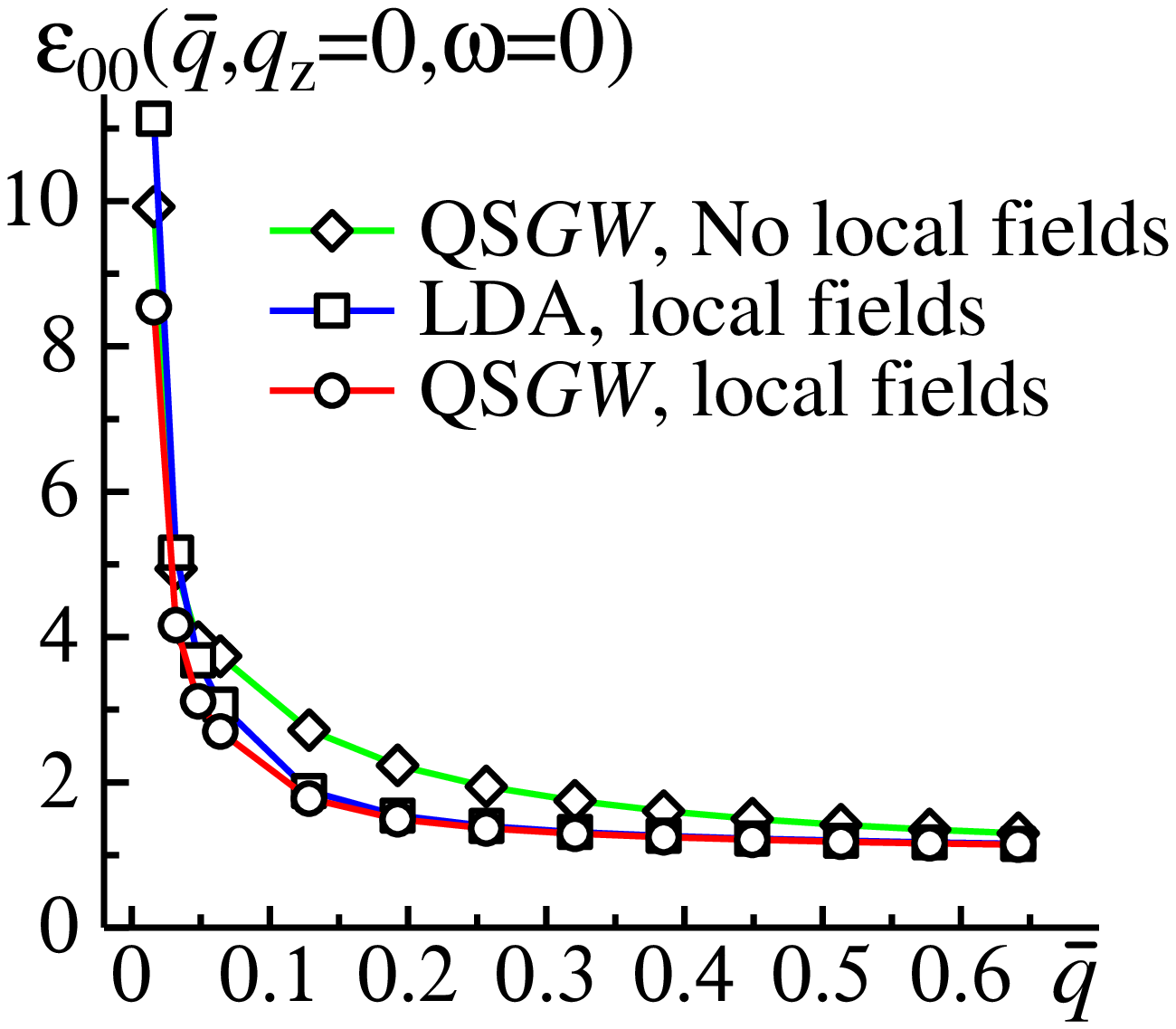}
\includegraphics[height=3.5cm]{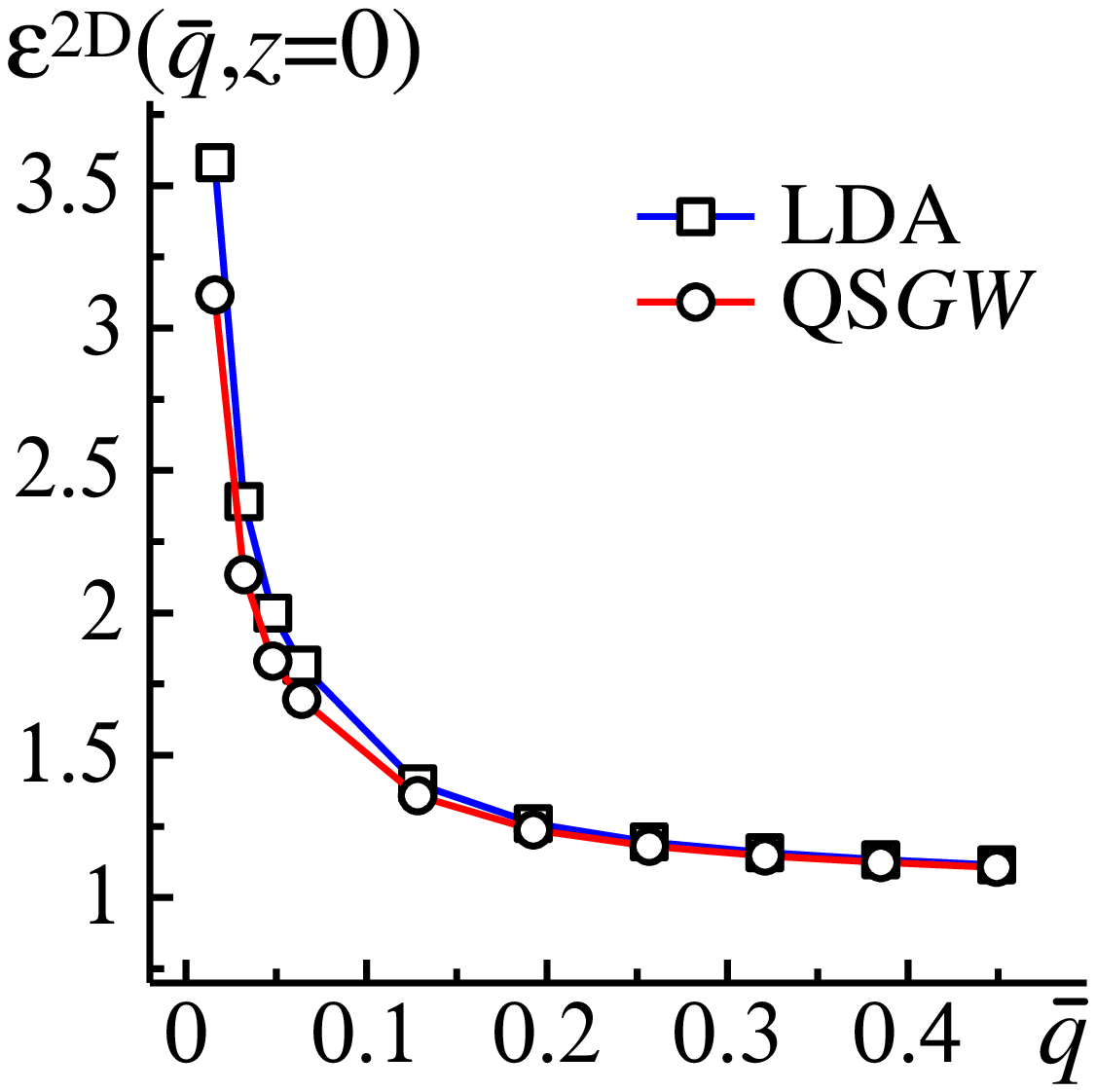}
\caption{(Top) Static dielectric function $\epsilon_{00}$($\bar{q},q_{z}$=0)
    along the (100) line in graphene, with local fields included and
    without.  $\bar{q}$ is in units of $2\pi/a$=2.56\AA$^{-1}$. The
    $q${}$\to$0 limit is delicate and there is some uncertainty in its
    value.  Shown for comparison is the same function calculated from the
    LDA potential.  In the $\bar{q}${}$\to$0 limit, $\epsilon_{00}$
    calculated by \qsgw\ is $\sim$0.8 smaller than the LDA result, similar
    to the ratio found in bulk semiconductors.  (Bottom) Effective layer
    dielectric function $\epsilon^{2{\rm D}}$($\bar{q},z$=0) as defined by
    Eq.~(\ref{eq:eps2d}), with local fields, calculated within the \qsgw\
    and LD approximations.  Local fields significantly reduce
    $\epsilon_{00}$.  The LDA result for $\epsilon_{00}$($\bar
    q$=0.086,$q_{z}$=0) without local fields is $\simeq$4, which
    agrees with the $\omega${$\to$}0 limit in Fig. 11 of
    Ref.\onlinecite{Marinopoulos04}.}
\label{fig:epsgraphene}
\end{figure}

$\epsilon_{00}$($\bar{q},q_z$) was calculated on a grid of points $\{\bar{q},q_z\}$;
the $q_z$=0 case is shown in the first panel of Fig.~\ref{fig:epsgraphene}.
It was found that $\epsilon_{00}$ is well parametrized (max error $<$0.1) by 
\begin{eqnarray}
\label{eq:epsfit}
\epsilon_{00}^{-1} (\bar{q},q_z ) &=& \frac{{a^2 (\bar{q}) + q_z^2 }}
                                          {{\epsilon_{00} (\bar{q},0)\,a^2 (\bar{q}) + q_z^2 }}\\
\label{eq:fita}
a^2 (\bar{q}) &=& \frac{{a_0 a_1 \bar{q}^2 }}{{a_1  + \bar{q}^2 }} \approx a_0 \bar{q}^2
\end{eqnarray}
where $a_0$=1.3 and 1.2 for \qsgw\ and LDA, respectively, and $a_1$=1.6$(2\pi/a)^2$.
The approximate form for $a$ in Eq.~\ref{eq:fita} is sufficient for any
$q$ where $\epsilon_{00}$ differs significantly from unity.  With
Eq.~(\ref{eq:epsfit}) $W^{2{\rm{D}}}$ can be integrated analytically.
Taking the approximate expression for $a^2(\bar{q})$ we obtain
\begin{eqnarray}
\epsilon ^{2{\rm{D}}} (\bar{q},z) = \frac{{\gamma (\gamma^2  - 1)}}{{\gamma (a_0  - 1) + (\gamma^2  - a_0 )e^{(1 - \gamma )\bar{q}z} }}
\label{eq:}
\end{eqnarray}
where $\gamma = \sqrt {a_0 \epsilon_{00}(\bar{q},0)}$.

Fig.~\ref{fig:epsgraphene} shows both kinds of dielectric functions,
$\epsilon_M$ corresponding to the macroscopic polarizability, and the
effective 2D static dielectric function
$\epsilon^{2{\rm{D}}}$($\bar{q},z$=0) calculated from
Eq.~(\ref{eq:epsfit}). Local fields reduce the strength of the screening.
The difference between LDA and \qsgw\ results are modest; and as noted
earlier, the LDA results are likely to be slightly better because they
benefit from a fortuitous cancellation of errors.  As $\bar{q}${}$\to$0,
$\gamma$ is significantly larger than $a_0$ and unity.  Keeping only the
leading order in $\gamma$, we obtain the surprising result that
$\epsilon^{2{\rm{D}}}$(0,$z$=0){}$\approx${}$\sqrt{a_0\epsilon_{00}(\bar{q},q_z\hbox{=}0)}$.
$\epsilon^{2{\rm{D}}}$(0,$z$=0) is roughly a factor 1.5 times larger than
the Dirac Hamiltonian result at zero doping.  Such a model predicts
$\epsilon(q) \approx 2.4$ independent of $q$, as shown by
Ando~\cite{ando_jpsj_2006}.  We find
$\epsilon^{2{\rm{D}}}$($\bar{q},z$=0){}$\approx$3.5 for $\bar{q}${}$\to$0,
but $\epsilon^{2{\rm{D}}}$ is a very strong function of $\bar{q}$.

Although virtual transitions involving Van Hove peaks of the density of
states might strongly enhance \cite{KT_anom,KT_anomb}
$\epsilon^{2{\rm{D}}}$ were they sufficiently close to the Fermi level,
apparently lie too far away in graphene.  The case of small $\bar{q}$ ($\bar
q${}$\sim${}$k_F${}$\leq$10$^{-2}$~\AA$^{-1}$) is relevant for transport
properties.  In this region our first-principles calculations do not
dramatically contradict predictions of the Dirac model.  At the same time,
for the problem of supercritical Coulomb centers and relativistic collapse
(fall on the center) \cite{shytov_prl_2007,shytov2,pereira_prl_2007}
distances of order of several inverse lattice constants are essential (this
is the radius of screening cloud, according to renormalization group
analysis \cite{shytov_prl_2007}), which corresponds to larger $q$.  For this
region our results show that the Dirac model {\it overestimates} the
screening.

\begin{acknowledgments}
MIK acknowledges support from Stichting voor Fundamenteel
Onderzoek der Materie (FOM), the Netherlands.  MvS was supported by
ONR contract N00014-7-1-0479 and NSF QMHP-0802216.

\end{acknowledgments}

\vfil\eject


\begin{thebibliography}{40}
\bibitem[{bas()}]{basis}
\bibinfo{note}{To ensure completeness in the interstital floating orbitals were
  placed in the ``empty sites'' where nucleii would fall if graphene were
  turned nto graphite. An $spdfspd$ basis was used for C atoms.}
\end{thebibliography}

\end{document}